\documentclass[RNAAS]{aastex631}

\shorttitle{LOFAR non-detections of SN\,2023ixf}
\shortauthors{Timmerman et al.}

\begin{document}

\title{LOFAR non-detections of SN\,2023ixf in its first year post-explosion}

\author{R.\ Timmerman} 
\affiliation{Centre for Extragalactic Astronomy, Department of Physics, Durham University, Durham, DH1 3LE, UK}
\affiliation{Institute for Computational Cosmology, Department of Physics, Durham University, South Road, Durham DH1 3LE, UK}

\author{M. Arias} 
\affiliation{ASTRON - Netherlands Institute for Radio Astronomy, Oude Hoogeveensedijk 4, 7991\,PD Dwingeloo, The Netherlands}

\author{A. Botteon} 
\affiliation{INAF - Instituto di Radioastronomia, Via P. Gobetti 101, 40129, Bologna, Italy}

\begin{abstract}

We used the LOFAR telescope to monitor SN\,2023ixf, a core-collapse supernova in M101, between 8 and 368 days post-explosion. We report non-detections down to $\sim80~\mu$Jy sensitivity at 144~MHz. Our non-detections are consistent with published radio observations at higher frequencies. At the time, we are not able to constrain the properties of low-frequency absorption due to the progenitor star's circumstellar medium via these LOFAR observations. 

\end{abstract}


\section{Introduction} \label{sec:intro}

Supernovae (SNe) emit radiation across the electromagnetic spectrum. Radio observations of SN light curves are complementary to those at infrared, optical, UV, and X-ray frequencies, which are sensitive to thermal emission
\cite[e.g.,][]{dwek81,rabinak11}. 
The radio emission has a non-thermal origin as it probes the SN shock's interaction with its environment \citep{chandra17}; however, there can be significant absorption, either synchrotron self-absorption from the emitting plasma, or free-free absorption due to cold foreground material. 

The core-collapse, type IIP supernova SN\,2023ixf discovered by \citet{itagaki2023} in Messier~101 \citep[d=6.85 Mpc;][]{riess22} on 19 May 2023 marked the closest supernova explosion since SN\,2014J, nine years prior \citep{fossey14}. Crucially, since SN\,2014J, the international stations of the LOw Frequency ARray \citep[LOFAR;][]{haarlem13} have become fully operational. This means that SN\,2023ixf presents a unique opportunity to study the low-frequency properties of radio supernovae.

SN~2023ixf has been proposed to have an abundant, dense circumstellar medium (CSM) ejected in the decades prior to the explosion
\citep[e.g.,][]{jacobson-galan23},
which will obscure its radio emission. A monitoring of the source in time, as it becomes progressively brighter at low radio frequencies, can help determine the properties of this CSM.

Here, we report on the results of a monitoring campaign performed with LOFAR at a frequency of 144~MHz, starting on 26 May 2023 and lasting up to the shutdown of LOFAR at the end of May 2024 for its scheduled upgrade. We made use of the full international array in order to simultaneously increase the effective collecting area of LOFAR as well as to help resolve potential radio emission associated with SN\,2023ixf from the diffuse emission of Messier~101.

\section{Observations \& Calibration}

A total of 14 observation epochs were obtained with LOFAR on SN\,2023ixf (14\(^\mathrm{h}\)03\(^\mathrm{m}\)38.56\(^\mathrm{s}\), 54\(^\circ\)18$'$41.9$''$) using Director's Discretionary Time (Project codes: DDT19\_005, DDT20\_002, DDT20\_008; PI: Arias). All observations consisted of 2~hours of on-target integration time as well as a 10-minute calibration scan before and after the target observation. Data were recorded between 120 and 168~MHz with an integration interval of 1~second and a frequency channel width of 3~kHz.

The data were calibrated per observation using the \textsc{Prefactor} software package \citep{vanweeren16, williams16, gasperin19}. We adopted the same calibration strategy for the International LOFAR Telescope as used in \citet{timmerman22}.
The calibrated data were imaged using \textsc{WSClean} \citep{offringa14} to search for emission associated with SN\,2023ixf.

\section{Results \& Discussion}

\begin{figure}
\plotone{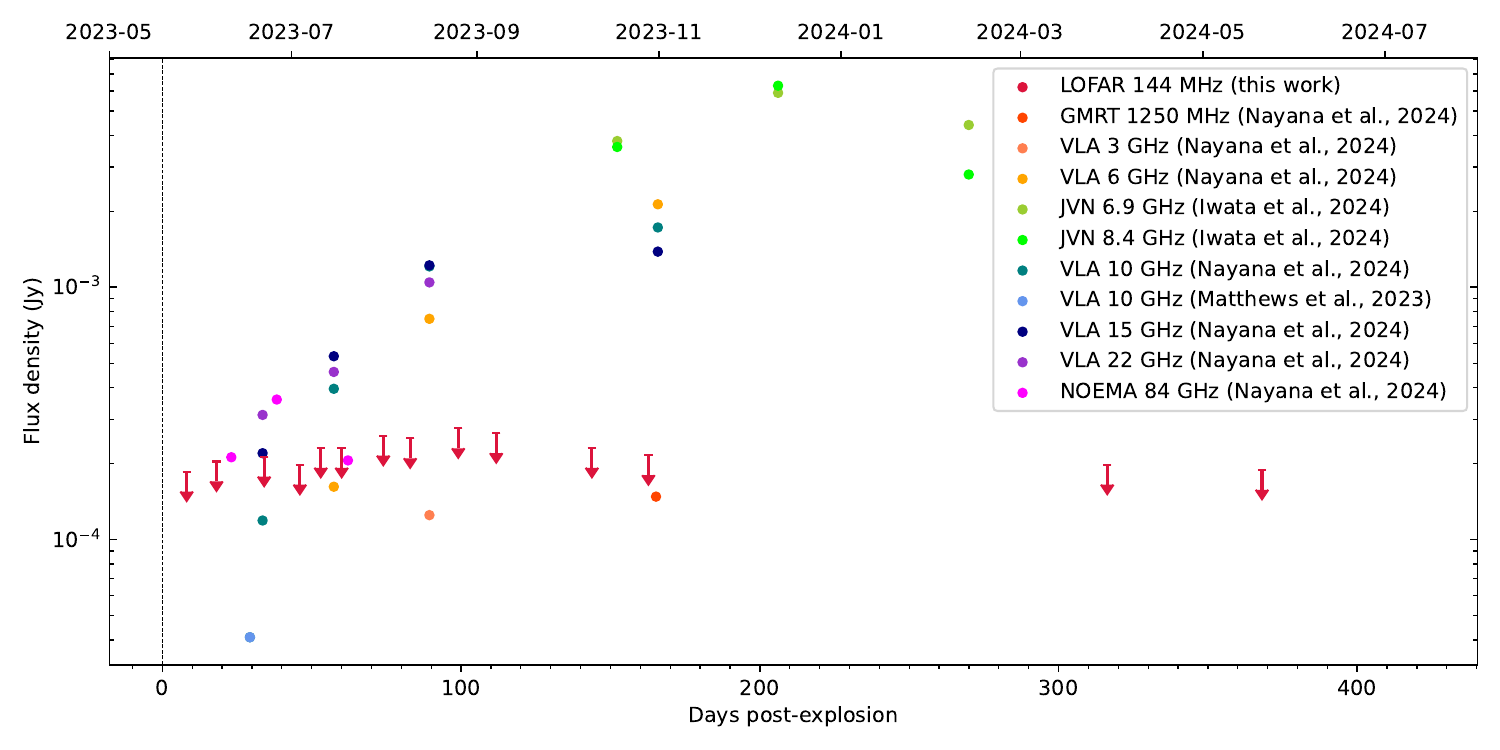}
\caption{Radio continuum detections of SN\,2023ixf obtained at higher frequencies combined with the non-detections obtained from LOFAR observations indicated at the \(3\sigma\) noise level.\label{fig:radiolightcurve}}
\end{figure}

We searched the final images for radio emission in the direction of SN\,2023ixf but were unable to detect significant emission in any of our epochs. Our non-detections are plotted in Figure~\ref{fig:radiolightcurve} as a function of days since estimated first light on 18 May 2023 at 17:50 UT \citep{hiramatsu23}.

\nocite{iwata24}\nocite{nayana24}\nocite{matthews24}

\cite{nayana24} present radio observations of SN~2023ixf taken $4-165$~days post-explosion with the Very Large Array (VLA), the Giant Meterwave Radio Telescope (GMRT), and the Northern Extended Millimeter Array (NOEMA). The SN radio light curve rises significantly later than for other SNe~II-P. The authors conclude that the radio emission is synchrotron, modulated by the free-free absorption caused by a dense CSM, likely the result of enhanced stellar mass loss in the $\sim100$~years preceding the SN explosion.

\cite{nayana24} first model the radio emission from SN~2023ixf as a smooth broken power law, with an optically thin and an optically thick term. This assumes that the emission is synchrotron radiation, and the absorption is synchrotron self-absorption. The presence of the free-free absorbing term would further dim the radio emission at lower frequencies.
Our non-detections are consistent with the best-fit parameters they derive for the synchrotron-only model from their data ($t \leq 165$~days). \cite{nayana24} also fit the turnover frequency from the optically thin to the optically thick regime, and the peak flux density as power-law functions of time. Using these functions, the predicted flux density at 150~MHz on day 400 post-explosion is $\sim14~\mu$Jy, still under our detection threshold. This means that at this time our observations cannot detect the synchrotron radiation, let alone constrain the parameters of a free-free absorbing component.
With LOFAR 2.0 set to begin science operations in early 2026 (roughly two and a half years post-explosion), the synchrotron flux at LOFAR frequencies should be within our detection limit, at approximately $\sim200~\mu$Jy.

Early-time low-frequency ($<200$~MHz) radio observations of extragalactic supernovae were not possible before the advent of LOFAR. None has been detected to date, since sufficiently nearby radio SNe are rare, and, as we have seen with SN~2023ixf, absorption effects are important at these early times. However, the rise time of SN~2023ixf in the radio has been unusually slow; for future SNe a slower initial cadence, perhaps of a couple of months, would constitute a better monitoring strategy.

\begin{acknowledgments}
\small
We thank the director of the International LOFAR Telescope (ILT) for granting us Director's Discretionary Time. 
RT is grateful for support from the UKRI Future Leaders Fellowship (grant MR/T042842/1). This work was supported by the STFC [grants ST/T000244/1, ST/V002406/1].  MA acknowledges support from the VENI research programme with project number 202.143, which is financed by the Netherlands Organisation for Scientific Research (NWO). 
This paper is based on data obtained with the LOFAR telescope (LOFAR-ERIC) under project codes DDT19\_005, DDT20\_002 and DDT20\_008. LOFAR (van Haarlem et al. 2013) is the Low Frequency Array designed and constructed by ASTRON. It has observing, data processing, and data storage facilities in several countries, that are owned by various parties (each with their own funding sources), and that are collectively operated by the LOFAR European Research Infrastructure Consortium (LOFAR-ERIC) under a joint scientific policy. The LOFAR-ERIC resources have benefited from the following recent major funding sources: CNRS-INSU, Observatoire de Paris and Université d'Orléans, France; BMBF, MIWF-NRW, MPG, Germany; Science Foundation Ireland (SFI), Department of Business, Enterprise and Innovation (DBEI), Ireland; NWO, The Netherlands; The Science and Technology Facilities Council, UK; Ministry of Science and Higher Education, Poland.
\end{acknowledgments}

\bibliography{references}{}
\bibliographystyle{aasjournal}

\end{document}